

\def\simle{{}^<_{\sim}}
\def\simge{{}^>_{\sim}}
\magnification = 1200
\baselineskip 14pt plus 2pt
\bigskip

\centerline{\bf ABSENCE OF A LOWER LIMIT ON $\Omega_b$ IN}
\centerline{\bf INHOMOGENEOUS PRIMORDIAL NUCLEOSYNTHESIS}
\vskip 0.7in
\centerline{Karsten Jedamzik and Grant J. Mathews}
\centerline{Physics Research Program}
\centerline{Institute for Geophysics and Planetary Physics}
\centerline{University of California}
\centerline{Lawrence Livermore National Laboratory}
\centerline{Livermore, CA 94550}
\centerline{ }
\centerline{and}
\centerline{}
\centerline{George M. Fuller}
\centerline{Department of Physics}
\centerline{University of California, San Diego}
\centerline{La Jolla, CA 92093-0319}
\vskip 0.65in

\centerline{\bf ABSTRACT}
We show that a class of inhomogeneous big bang nucleosynthesis models exist
which yield light-element abundances in agreement with observational
constraints for baryon-to-photon ratios significantly smaller than those
inferred from standard homogeneous big bang nucleosynthesis (HBBN).
These inhomogeneous
nucleosynthesis models are characterized by a bimodal distribution of baryons
in which some regions have a local baryon-to-photon ratio $\eta\approx
3\times 10^{-10}$, while the remaining regions are baryon-depleted.
HBBN scenarios with primordial ${\rm (^2H+{}^3He)/H}\ \simle\  9\times 10^{-5}$
necessarily require that most baryons be in a dark or non-luminous form,
although new observations of a possible high deuterium abundance in
Lyman-$\alpha$
clouds may relax this requirement somewhat. The models described here
present another way to relax this requirement and can even  eliminate any
lower bound on the baryon-to-photon ratio.

\vskip 0.15in
{\it  Subject headings: early universe - abundances,  nuclear reactions,
nucleosynthesis,}
\hskip 0.5in
{\it cosmology - dark matter}\hfill\break

\vfill\eject
\baselineskip 14pt plus 2pt

\centerline{\bf 1. Introduction}
\vskip 0.1in

In this paper we point out a feature of
inhomogeneous primordial nucleosynthesis
scenarios which to our knowledge has not been previously emphasized. In
particular, we show that inhomogeneous big bang nucleosynthesis scenarios
(hereafter; IBBN) could lead to a relaxation of the lower limit on $\Omega_b$.
This may have important implications for the problem of the \lq\lq
missing\rq\rq\ or dark baryons. In what follows we briefly review the problem
of the missing baryons. We then discuss IBBN scenarios which have very low
$\Omega_b$, but which otherwise produce light-element abundance yields in
agreement with observation.

\vskip 0.15in
\centerline{\bf 1.1. Luminous Matter}
\vskip 0.1in

A lower bound on the baryonic contribution to the closure density, $\Omega_b$,
can be obtained from estimating the baryonic content of luminous objects. These
objects include spiral and elliptical galaxies, as well as X-ray emitting
diffuse intergalactic gas in groups and clusters of galaxies. Significant
amounts of cold hydrogen gas is also observed at high redshift in
Lyman-$\alpha$ clouds. If this gas lies in front of quasars it can be detected
through it's
absorption features (cf. Wolfe 1988).

The density of baryons in luminous objects can be simply estimated. It is
obtained by multiplying the observed luminosity density, $\cal L\ $, by a
typical \lq\lq mass-to-light ratio\rq\rq $(M/L)$ (in units of
mass-per-luminosity). The sum over spiral galaxies, elliptical galaxies, and
diffuse intergalactic gas then yields the ratio of the baryon density in
luminous objects, $\rho_b^{lum}$, to the closure density, $\rho_c$:
$$\Omega_b^{lum}={\rho_b^{lum}\over \rho_c}={1\over\rho_c}\sum_i {\cal L}_i
\biggl({M\over L}\biggr)_i\ .\eqno(1)$$
Contributions from Lyman-$\alpha$ clouds are often excluded from the sum in
eq.(1). The rational for this exclusion is that it is not yet clear to what
extent baryons in Lyman-$\alpha$ clouds are eventually incorporated into
galaxies and intergalactic gas already accounted for in eq.(1).

The luminous baryon content of the universe has been estimated by a number of
authors (e.g. Peebles 1971; Gott {\it et al.} 1974; Olive {\it et al.} 1981;
B\"orner 1988; Hogan 1990; White 1990; Persic \& Salucci 1992). Most estimates
of $\Omega_b^{lum}$ fall in the interval
$$0.003 \simle\ \Omega_b^{lum}\ \simle 0.007\ .\eqno(2)$$
Uncertainties in these estimates reflect uncertainties in both the observed
luminosity densities ${\cal L}$ and the adopted mass-to-light ratios $(M/L)$.
Note that the inferred range for $\Omega_b^{lum}$ exhibits only a very weak
dependence on the Hubble constant.

Persic and Salucci (1993) estimate that the cosmic baryon density could be as
small as $\Omega_b^{lum}\approx 0.003$.
These authors argue that $\Omega_b^{lum}$ is smaller than previously estimated
by as much as a factor of two based upon an attempt
to account properly for the fact that mass-to-light ratios decline with
decreasing galaxy luminosity.
It is interesting to note that the estimate by Persic and Salucci is close to
that for the baryon density in Lyman-$\alpha$ clouds,
$\Omega_{Ly}\approx 0.002-0.003$ (Wolfe 1988; Lanzetta {\it et al.} 1991).
In any case, there seems to be a consensus that the cosmic baryon density in
luminous objects can not be much larger than $\Omega_b^{lum}\approx 0.01$.
This conclusion is independent of the value of the Hubble constant.

\vskip 0.15in
\centerline{\bf 1.2. Standard Homogeneous Big Bang Nucleosynthesis}
\vskip 0.1in

Calculations of standard homogeneous big bang nucleosynthesis (hereafter; HBBN)
provide an independent prediction for the baryon content of the universe.
Observationally inferred light-element abundances of $^2$H, $^3$He, $^4$He, and
$^7$Li agree well with calculated primordial nucleosynthesis abundance yields,
whenever
$\Omega_b^{\rm HBBN}$ is in a small range of values centered around
$\Omega_b^{\rm HBBN}\approx 0.046h_{50}^{-2} (T_{2.75})^{3}$
(Wagoner, Fowler, \& Hoyle 1967;
Wagoner 1973; Schramm \& Wagoner 1977; Yang {\it et al.} 1984; Krauss \&
Romanelli
1990; Walker {\it et al.} 1991, Smith, Kawano, \& Malaney 1993) where
$h_{50}$ is the Hubble constant in units of 50 km s$^{-1}$ Mpc$^{-1}$, and
$T_{2.75}$ is the present microwave background temperature in units
of 2.75 K. When computational,
observational, and nuclear reaction rate uncertainties are taken into account,
the allowed range for $\Omega_b^{\rm HBBN}$ is (Smith {\it et al.} 1993)
$$0.043 {{}^<_{\sim}}\ \Omega_b^{\rm HBBN}h_{50}^{2}(T_{2.75})^{-3}\
{{}^<_{\sim}} 0.056\ .\eqno(3)$$
Here the lower limit on $\Omega_b^{\rm HBBN}$ arises mainly from deuterium
overproduction. Current estimates of the Hubble constant range between
$0.8 \simle h_{50} \simle 1.7$ (cf. van den Bergh 1989).  The present best
determination of the microwave background temperature from the {\it COBE}
satellite is 2.726 K $\pm$ 0.010 ($T_{2.75}=0.9912\pm 0.0036$) (Mather {\it et
al.} 1994). The weighted
mean of the {\it COBE} measurement with others at
wavelengths greater than 1 mm is 2.76 $\pm$ 0.10 ($T_{2.75}=1.004\pm 0.004$)
(Smith {\it et al.} 1993).
In what follows we will omit the dependence of $\Omega_b$ on the rather
accurately known CMBR-temperature.

It is clear upon comparison of eq.(2) and eq.(3) and from considerations of the
value of the Hubble constant, that the baryon density predicted by HBBN is
likely to exceed the baryon density inferred from luminous objects by a factor
possibly as large as 10.
This would require the bulk of baryons in the universe to be dark.
A vexing question in the standard model of cosmology is how most of the baryons
come to be in a non-luminous form.

Recently, Songaila {\it et al.} (1994) have reported detection of an
isotope-shifted Lyman-$\alpha$ deuterium absorption line at high redshift along
the line of sight to a quasar. They report a deuterium abundance of $1.9\times
10^{-4}\simle {\rm ({}^2H/H)}\simle 2.5\times 10^{-4}$. If this value is
interpreted as a primordial abundance than it
is significantly larger than the previously accepted upper limit on this
quantity, $({\rm ^2H+{}^3He}/H)\simle 9\times 10^{-5}$ (Smith {\it et al.}
1993;
Walker {\it et al.} 1991). It is not yet clear whether the new number for
${\rm (^2H/H)}$ should be accepted as the primordial abundance, since the
probability
of a systematic error from a Lyman-$\alpha$ absorber could be large.

If we take the primordial deuterium abundance to be $1.9\times 10^{-4}\simle
{\rm ({}^2H/H)}\simle 2.5\times 10^{-4}$ then the range of $\Omega_b$ inferred
from HBBN changes to
$$0.022\simle\ \Omega_b^{\rm HBBN}h_{50}^{2}\ \simle 0.026\ .\eqno(4)$$
These values of $\Omega_b^{\rm HBBN}$ could be reconciled with $\Omega_b^{lum}$
without demanding that most baryons be dark, so long as the Hubble
parameter is large. Note that in this case, however, there may be uncomfortably
little room for any dark baryons if $\Omega_b^{lum}$ is near the upper end of
its
observationally inferred range. In this extreme case
the kind of inhomogeneities we discuss in this paper are constrained.

\vskip 0.15in
\centerline{\bf 1.3. Dark Baryons}
\vskip 0.1in

Several ways of hiding baryons in dark objects have been suggested. However,
most of these scenarios have potential drawbacks or can be ruled out by
observation. In view of the complexity of the dark matter problem we will not
present a complete discussion here, but rather refer the reader to recent
review articles on the subject (Trimble 1987; Hogan 1990; Ashman 1992). Two
potential sites for non-luminous baryons are: 1) a smooth intergalactic ionized
background of baryons which is not incorporated into galaxies at the present
epoch; and 2) compact objects in galactic halos such as planets, brown dwarfs,
white dwarfs, or black holes.
An intergalactic baryonic component could in principle account for the missing
baryons, but this gas would have to be ionized. If the gas were ionized then it
would not be detectable by absorption features in the spectrum of distant
galaxies and quasars. However, the temperature of the gas could not exceed
$T\sim 10^8$ K or its X-ray emission would be observable (Peebles 1971).

It is unclear whether compact objects in the halo which may account for the
missing baryons could be comprised principally of low-mass stars. The
uncertainty
is due to a lack of reliable estimates of the luminosity density from such
objects (cf. Richstone {\it et al} 1992; Burrows 1994). In principle, white
dwarfs could exist in large numbers in the halo without having been detected.
However, this would imply that the initial mass function (IMF) was strongly
peaked around $4M_{\odot}$. If the IMF were not strongly peaked around this
mass too many low-mass stars and/or neutron stars would be produced
(Ryu {\it et al.} 1990). The progenitors of neutron stars would produce
heavy elements. Large numbers of neutron stars in
the halo might lead to overproduction of heavy elements at an early epoch in
the history of the galaxy.

Probably the best candidates for baryonic compact objects in the halo are
brown dwarfs with masses $M\ {}^<_{\sim}\ 0.008M_{\odot}$ and/or massive black
holes with masses $M\ {}^>_{\sim}\ 200 M_{\odot}$ (Carr {\it et al.} 1984; Carr
1990). Here, black holes count as baryonic dark matter only if they
predominantly were formed from baryons and their formation occurred after the
epoch of primordial nucleosynthesis. These black holes could not exceed a mass
of about $M\approx 10^{6.5}M_{\odot}$ or structures associated with galactic
disks would be disrupted (Lacey \& Ostriker 1985).

An abundant brown dwarf population requires a sharp increase in the IMF at or
below the hydrogen burning limit, $M\approx 0.08M_{\odot}$. This requirement
stems from the desire not to overproduce low-mass, hydrogen-burning stars. In
any case, a star formation process which is intrinsically different from that
seen in current star formation regions would be required in order for either
brown dwarfs or black holes to be the hiding places for non-luminous baryons.

The recent results of gravitational micro-lensing experiments (Alcock {\it
et al.} 1993; Aubourg {\it et al.} 1993) may indicate that at least some
component of galactic halo dark matter is comprised of condensed objects.
However, these experiments are not definitive as to the composition of
these objects. For example, these objects may be low-mass baryonic stars or
brown dwarfs, but conceivably these objects could be primordial black holes,
topological defects, or mass-energy in some other form which does not
(or did not) carry significant net baryon number. It seems likely to us,
however, that these objects are baryonic. If this turns out to be the case,
then astrophysicists are faced with the problem of how baryons get into such a
low-mass condensed state without violating constraints on galactic
chemical evolution and dynamics. If in the future it is determined that the
gravitational microlensing objects are either non-baryonic or that baryonic
micro-lensing objects constitute only a small fraction of the halo mass, then
the question of where the baryons are hidden and our speculations
on the role of the IBBN models and the lower limit on $\Omega_b$ becomes
relevant.

If the future gravitational microlensing observations infer that there is a
dark matter content equivalent to $\Omega^{Halo}\approx 0.03 - 0.07$, then
there may be a problem in interpreting this dark matter as baryonic in origin
if the primordial deuterium abundance satisfies $1.9\times 10^{-4}\simle ({\rm
^2H/H}) \simle 2.5\times 10^{-4}$. In this case, we could conclude that either
the objects are not baryonic or the primordial nucleosynthesis process has been
influenced significantly by density fluctuations (Gnedin \& Ostriker 1992, Cen,
Ostriker, \& Peebles 1993, Jedamzik \& Fuller 1994).
\vfill\eject
\centerline{\bf 2. Baryon Inhomogeneous Big Bang Nucleosynthesis}
\vskip 0.1in

Inhomogeneous big bang nucleosynthesis scenarios were motivated originally by
Witten's speculations about a first-order cosmic QCD-phase transition and it's
effects on the cosmic distribution of baryon number (Witten 1984). Subsequent
work on IBBN models has addressed the question of whether there is a way around
the HBBN upper limit on $\Omega_b$ (Alcock, Fuller, \& Mathews 1987; Applegate,
Hogan, \& Scherrer 1987; 1988; Fuller, Mathews, \& Alcock 1988; Kurki-Suonio
{\it et al.} 1988; 1990; Malaney \& Fowler 1988; Boyd \& Kajino 1989; Terasawa
\& Sato 1989abc; 1990; Kajino \& Boyd 1990; Kurki-Suonio \& Matzner 1989; 1990;
Mathews {\it et al.} 1990; 1993; Kawano {\it et al.} 1991; Jedamzik, Fuller, \&
Mathews 1994; Thomas {\it et al.} 1994).
Most recently it has been shown (e.g. Jedamzik {\it et al.} 1994) that for
spherically condensed fluctuations the upper limit on $\Omega_b$ is virtually
unchanged when compared to the upper limit on $\Omega_b$ derived from HBBN.

In the present paper, however,  we wish to point out that in inhomogeneous
nucleosynthesis scenarios at low average baryon-to-photon ratio
(corresponding to $\Omega_b<0.046h_{50}^{-2}$) fluctuations with the
right characteristics can yield primordial light-element abundances which agree
with
observationally inferred limits. Given the right fluctuation characteristics
there
is essentially no {\it lower} limit on $\Omega_b$.

The type of fluctuation in a low average $\Omega_b$ universe which shows
agreement between calculated light-element abundances and observationally
inferred abundance limits is shown schematically in Fig. 1.
In this figure we show the distribution of baryon-to-photon ratio $\eta$ as a
function of length scale $x$. The universe is seen to be made up of two
distinct environments: 1) high-density regions with local $\eta^h\approx
3\times 10^{-10}$; and 2) low-density regions with local baryon-to-photon ratio
$\eta^l<< 3\times 10^{-10}$, so that low-density regions are essentially
evacuated of baryons. Agreement between calculated light-element
nucleosynthesis yields and observationally inferred abundance limits is
attained in these models because the high-density regions have $\eta^h\approx
3\times 10^{-10}$ (corresponding to $\Omega_b^h\approx 0.046h_{50}^{-2}$) which
is the preferred baryon-to-photon ratio in HBBN.  Local abundance yields in
high-density regions are then indistinguishable from abundance yields
resulting from HBBN.
Abundance yields averaged over high- and low-density regions will be
indistinguishable
from abundance yields in HBBN if the fraction of baryons
residing in low-density
regions is much smaller than the fraction of baryons residing in high-density
regions.

Note that in such IBBN scenarios the averaged baryon density, or equivalently
${\bar\Omega_b}$, will be smaller than the preferred HBBN value.
Assuming that a volume fraction $f_V$ of the universe is at $\eta\approx
3\times 10^{-10}$, and approximating the remaining volume fraction $(1-f_V)$ to
be evacuated of baryons, we infer an average baryon density ${\bar\Omega_b}$
$${\bar\Omega_b}\approx\Omega^{\rm HBBN}_b f_V \approx 0.046h_{50}^{-2} f_V\
,\eqno(5)$$
a value which can be much smaller than $\Omega_b^{\rm HBBN}\approx
0.046h_{50}^{-2}$.

\vskip 0.15in
\centerline{\bf 2.1. Constraints from Baryon Diffusion}
\vskip 0.1in

Of course, abundance yields resulting from an inhomogeneous baryon
distribution,
such as that shown in Fig. 1, can only match abundance yields of standard
homogeneous
primordial nucleosynthesis if the effects of diffusive and hydrodynamic damping
processes on fluctuations during the nucleosynthesis era are negligible.
This requirement implies that the average mean separation between fluctuation
sites $l$ should exceed neutron-, proton-, and photon- diffusion lengths
during the epoch of primordial nucleosynthesis. We have calculated the
abundance
yields of spherically condensed fluctuations with step-function profiles,
similar
to the fluctuations shown in Fig. 1, as a function of fluctuation separation
distance $l$. For this calculation we have assumed a regular lattice of
fluctuation sites.
We have fixed the baryon-to-photon ratio in the spherical high-density regions
at $\eta^h=3.1\times 10^{-10}$ and the baryon-to-photon ratio in the
low-density
regions at $\eta^l=3.1\times 10^{-15}$. By assuming a volume fraction
$f_V=0.065$
of the universe to be at high baryon-to-photon ratio, we fix the average
$\Omega_b$ in our model at ${\bar\Omega_b}=0.003h_{50}^{-2}$ in agreement with
the lower limit on $\Omega_b^{lum}$. In Fig. 2 we show the calculated abundance
yields for $^2$H plus $^3$He, $^4$He, and $^7$Li resulting from such
fluctuations as a function of separation distance between adjacent fluctuation
sites $l_{100}$. Here $l_{100}$ is the proper fluctuation separation distance
at an epoch where the cosmic temperature is $T=100$ MeV. It is evident from the
figure that for $l_{100} \simge 10^4$m abundance yields in our model with
${\bar\Omega_b}=0.003h_{50}^{-2}$ are indistinguishable from the abundance
yields of a homogeneous
primordial nucleosynthesis scenario with $\Omega_b=0.046h_{50}^{-2}$.

For values of $l_{100}$ smaller than $l_{100}\approx 10^4$m, deuterium
production
increases and $^4$He production decreases. This results from neutron diffusion
effecting a transfer from the high-density region to the low-density region. In
turn,
 this diffusive transport leads to the formation of extended transition regions
between high- and low-density regimes. The result may be a non-negligible
fraction
of baryons at low baryon-to-photon ratio and concomitant overproduction of
deuterium.
Deuterium yields increase rapidly with decreasing baryon-to-photon ratio.

It is therefore necessary that the separation of high-density regions
exceed $l_{100}\simge 10^4$m in order that deuterium overproduction be avoided.
The value of this lower limit on $l_{100}$ may be slightly increased if
other fluctuation geometries are considered. Examples of such alternative
 geometries include high-density spherical shells. A characteristic
baryonic mass content can be
assigned to the fluctuation cells.
For a fluctuation cell of radius $l_{100}\simge 10^4$m we find that the
baryonic mass
within each high-density region must exceed
$$M_b\ \simge\ 10^{-11}M_{\odot}\biggl({l_{100}\over 10^4{\rm
m}}\biggr)^3\biggl({
\Omega_bh_{50}^{2}\over 0.003}\biggr)\ ,\eqno(6)$$
in order to avoid deuterium overproduction.

An upper limit on the baryonic mass of such fluctuations can be obtained from
considerations of the small-scale isotropy of the cosmic microwave background
radiation (CBR). It is known that the anisotropies in the CBR on small angular
scales of 1-10 arcmin do not exceed $\Delta T/T\ \simle\ 5\times 10^{-5}$
(Readhead {\it et al.} 1989). A fluctuation at baryon-to-photon ratio
$\eta\approx 3\times 10^{-10}$ subtending
an angular scale of 1 arcmin at decoupling will contain approximately a
baryonic mass of $M_b\approx 10^{11}M_{\odot}$. Such large fluctuations will
maintain an increased internal temperature so that the fluctuation's
self-gravity is counterbalanced by the radiation overpressure. In order for the
resulting distortions in the CBR not to exceed the upper limit of $\Delta\
T/T\simle\ 5\times 10^{-5}$ on arcminute scales the baryonic mass within a
fluctuation cell has to be less than
$$M_b\ \simle\ 10^{11}M_{\odot}\ .\eqno(7)$$
Note that this mass limit is roughly the baryonic mass of a typical galaxy
and is many orders of magnitude above the lower limit given in eq.(6).

Deuterium overproduction also can be employed to place limits on the
fraction of baryons contained in the low-density regions. Likewise, the
fraction of baryons residing in transition regions between high- and
low-density regimes can be constrained. The total deuterium yield resulting
from a bimodal distribution such as the one displayed in Fig. 1
(i.e., a distribution without any transition region) is approximately
$$\bar{\biggl({{\rm D}\over {\rm H}}\biggr)}\approx \biggl({{\rm D}\over {\rm
H}}\biggr)_h + f_l\biggl({{\rm D}\over {\rm H}}\biggr)_l\ ,\eqno(8)$$
where $f_l$ is the fraction of baryons contained in the low-density regions,
and $({\rm D}/{\rm H})_h$ and $({\rm D}/{\rm H})_l$ are the local
deuterium-to-hydrogen number fractions in high-density and low-density regions,
respectively. In writing eq.(8) we have implicitly assumed that effects of
neutron diffusion during the nucleosynthesis era are negligible and that the
fraction of baryons residing in the low-density regions is small,
$f_l<<1$. The deuterium yield increases at lower baryon-to-photon ratio
from $({{\rm D}/{\rm H}})\approx 5.5\times 10^{-3}$ at $\eta =10^{-11}$ to
a maximum yield of $({\rm D}/{\rm H})\approx 9\times 10^{-3}$ at
$\eta =2\times 10^{-12}$ and then decreases to $({\rm D}/{\rm H})\approx
10^{-3}$ for $\eta =10^{-13}$. Thus, even a small fraction of baryons residing
in the low-density regions could make a significant contribution to the total
deuterium abundance. If we require the contribution to the deuterium yield
arising from the low-density regions not to exceed
$f_l({\rm D}/{\rm H})_l\ \simle\ 10^{-5}$, and assume deuterium production
in the low-density region to be at a level of $({\rm D}/{\rm H})_l\approx
10^{-3}$, we can obtain an upper limit on the fraction of baryons allowed to
reside in the low-density regions, $f_l \simle 0.01$.  For a universe with
$\eta_h = 3.1 \times 10^{-10}$, $f_v = 0.065$, and $\bar \Omega_b =
0.003 h_{50}^{-2}$ as above, this would imply that the baryon-to-photon ratio
in
the low density region should not exceed $\eta_l \simle 10^{-13}$.  In a
similar
way the fraction of baryons within transition regions can be constrained to be
smaller than $f_l \simle 0.01 - 0.001$.

The reader might conclude at this point that it is not surprising that
light-element nucleosynthesis can be made to agree with observation for a given
$\Omega_b$ because there are many adjustable parameters in IBBN models.
However, detailed numerical hydrodynamic studies of IBBN scenarios (cf.
Jedamzik {\it et al.} 1994) show how remarkably difficult it is to obtain
agreement with observation for baryon-to-photon ratios which substantially
deviate from $\eta\approx 3\times 10^{-10}$. However, even though
observationally inferred primordial abundance constraints demand that almost
all baryons must freeze out of nuclear statistical equilibrium with
$\eta\approx 3\times 10^{-10}$, these same constraints do not limit the
fraction of space that is filled by baryons.

Finally, we note that, even for a homogeneous distribution of baryons at cosmic
temperature $T\approx 100$ keV, the inferred
$\Omega_b$ can conceivably be lower
than that deduced from a standard cosmic scenario. This can be the case if,
after
a standard HBBN scenario with $\eta\approx 3\times 10^{-10}$, a large amount of
entropy is released into the CBR. Such a release of entropy
could result in a prolonged ionization or reionization of the universe and
would reset the ultimate baryon-to-photon ratio to a lower value. Possible
sources of significant entropy production after the epoch of primordial
nucleosynthesis could be an abundant primordial black hole population which
evaporated well before the present epoch, late phase transitions, or the
accretion of matter on an abundant early population of massive black holes.
However, there should exist stringent constraints on such scenarios, since
the evaporation of primordial black holes and/or the accretion of matter
on massive black holes would result in the production of $\gamma$-rays,
which in turn might reprocess the nuclear abundances by
photo-disintegration (Carlson et al. 1990; Gnedin \& Ostriker 1992).
Furthermore, a significant release of entropy could distort the CBR such that
the resulting CBR-spectrum would deviate from a Plankian spectrum (Mather {\it
et al.} 1990; 1994).

\vskip 0.15in
\centerline{\bf 3. Conclusions}
\vskip 0.1in

We have shown that there exist IBBN models which agree with observations, but
for low values of $\Omega_b$. These models are constrained however.
In particular, a lower limit on the baryonic
mass of fluctuations of $M_b\ \simge\ 10^{-11}M_{\odot}$ implies that a
speculative inhomogeneous electroweak baryogenesis scenario can not form the
type of inhomogeneity considered here, as the baryonic mass contained within
the horizon
during the electroweak epoch is only $\sim 10^{-18}M_{\odot}$.  The baryonic
mass
within the horizon at the QCD-epoch, however, is roughly
$M_b^{QCD}\sim 10^{-9}M_{\odot}\ $, which is close to the lower limit on the
mass
of fluctuations in eq.(6). Only an unlikely first-order QCD phase transition
scenario
in which there are a few fluctuations (or nucleation sites) per horizon volume
could lead to the formation of a fluctuation with these characteristics. In
the framework of a standard early universe
scenario baryogenesis associated
with an inflationary epoch could, in principle, form fluctuations on the
desired spatial scales. Fluctuations would have to be formed with a bimodal
character, with high-density regions having little spread around the
baryon-to-photon
ratio $\eta\approx 3\times 10^{-10}$ and baryon-poor low-density regions.
Furthermore the transition regions between high- and low-density should
contain only a small fraction of the baryons.

In summary, we have identified and constrained inhomogeneous primordial
nucleosynthesis scenarios with abundance yields which agree with
observationally inferred abundance limits yet have $\Omega_b$ much lower than
the lower limit on this quantity from HBBN. These models assume the universe to
be filled with high-density regions with $\eta\approx3.1\times 10^{-10}$ and
low-density regions with $\eta\simle 10^{-13}$. A lower limit on $\Omega_b$ in
these models is completely absent. Such primordial nucleosynthesis scenarios
offer an alternative solution to the problem of the missing or dark baryons.

\vskip 0.15in
\centerline{\bf 4. Acknowledgements}
\vskip 0.1in

The authors wish to thank C. R. Alcock for useful conversations and helpful
suggestions. This work was supported in part by NSF Grant PHY91-21623. It was
also performed in part under the auspices of the US Department of Energy by the
Lawrence Livermore National Laboratory under contract number W-7405-ENG-48.

\vfill\eject

\centerline{ }
\centerline{\bf 5. References}
\vskip 0.1in

\item{ }Alcock, C.R., Fuller, G.M., \&  Mathews, G.J. 1987, ApJ, 320,  439
\item{ }Alcock, C. {\it et al.} 1993, Nature, 365, 621
\item{ }Applegate, J.H., Hogan, C.J., \&   Scherrer, R.J. 1987, Phys. Rev.,
D35, 1151
\item{ }Applegate, J.H., Hogan, C.J., \&   Scherrer, R.J. 1988, ApJ, 329, 592
\item{ }Ashman, K.M. 1992, Pub.of Astr.Soc. of Pac., 104, 1109
\item{ }Aubourg, E. {\it et al.} 1993, 365, 623
\item{ }B\"orner, G. 1988, {\it The Early Universe}, Springer-Verlag, Berlin.
\item{ }Boyd, R.N.\&  Kajino, T. 1989, ApJ, 336, L55.
\item{ }Carlson, E.D., Esmailzadeh, R., Hall, L.J., Hsu, S.D.H. 1990, Phys.
Rev. Lett., 65, 2225
\item{ }Burrows, A. 1994, ApJ submitted
\item{ }Carr, B.J., Bond, J.R., \& Arnett, W.D. 1984, ApJ, 277, 445
\item{ }Carr, B.J. 1990, Comm.in.Astroph.C, 14, 257
\item{ }Cen, R., Ostriker, J. P., \& Peebles, P. J. E. 1993, ApJ, 415, 423
\item{ }Dearborn, D.S.P., Schramm, D.N., \& Steigman, G. 1986, ApJ, 302, 35
\item{ }Fuller,  G.M.,  Mathews, G.J., \&  Alcock, C.R. 1988, Phys. Rev., D37,
1380
\item{ }Gnedin, N.Y. \& Ostriker, J.P. 1992, ApJ, 400, 1
\item{ }Gott, J.R., Gunn, J.E., Schramm, D.N., \& Tinsley, B.M. 1974, ApJ, 194,
543
\item{ }Hogan, C.J. 1990, in {\it Baryonic Dark Matter} pg 1, editors
Lynden-Bell, D. \& Gilmore G., Kluwer, Dordrecht
\item{ }Jedamzik, K., Fuller, G. M., \& Mathews, G, J, 1994, ApJ, 423, 50
\item{ }Jedamzik, K.\& Fuller, G. M. 1994, in preparation
\item{ }Kajino, T. \&  Boyd, R.N. 1990, ApJ, 359, 267
\item{ }Kawano, L.H., Fowler, W.A., Kavanagh,  R.W., \& Malaney, R.A., 1991,
ApJ, 372, 1
\item{ }Krauss, L.M. \& Romanelli, P. 1990, ApJ, 358, 47
\item{  }Kurki-Suonio, H., \&  Matzner, R. A. 1989, Phys. Rev. D39, 1046
\item{  }Kurki-Suonio, H., \&  Matzner, R. A. 1990, Phys. Rev. D42, 1047
\item{ }Kurki-Suonio, H., Matzner, R.A., Centrella, J., Rothman, T., \&
    Wilson, J.R. 1988, Phys. Rev., D38, 1091
\item{ }Kurki-Suonio, H., Matzner, R. A., Olive, K. A., \& Schramm, D. N. 1990,
ApJ, 353, 406
\item{ }Lacey, C.G., Ostriker, J.P. 1985, ApJ, 299, 633
\item{ }Lanzetta, K.M., Wolfe, A.M., Turnshek, D.A., Limin Lu, ... 1991, ApJS,
77, 1
\item{ }Malaney, R.A. \&  Fowler, W.A. 1988, ApJ,  333, 14
\item{ }Mather, J.C., Cheng, E.S., Eplee, R.E. Jr.,,Isaacman, R.B., and others
1990, ApJL, 354, L37
\item{ }Mather, J.C., et al. 1994, ApJ submitted.
\item{ }Mathews, G.J., Meyer, B.S., Alcock, C.R., \&  Fuller, G.M. 1990,  ApJ,
358, 36
\item{ }Mathews, G.J., Schramm, D.N., \& Meyer, B.S. 1993, ApJ, 404, 476
\item{ }Olive, K.A., Schramm, D.N., Steigman, G., Turner, M.S., \& Yang, J.
1981, ApJ, 246, 557
\item{ }Persic, M. \& Salucci, P. 1992, MNRAS, 258, 14r
\item{ }Readhead, A.C.S., Lawrence, C.R., Myers, S.T., Seargent, W.L.W.,
Hardebeck, H.E., Moffit, A.T. 1989, ApJ, 346, 566
\item{ }Richstone, D., Gould, A., Guhathakurta, P., \& Flynn, C. 1992, ApJ,
388, 354
\item{ }Ryu, D., Olive, K.A., \& Silk, J. 1990, ApJ, 353, 81
\item{ }Schramm, D.N. \&  Wagoner, R.V. 1977, Ann. Rev. Nucl. Part. Sci., 27,
37
\item{ }Smith, M. S., Kawano, L. H. \&  Malaney, R. A., 1993, ApJS, 85, 219
\item{ }Songaila, A., Cowie, L. L., Hogan, C. J., \& Rugers, M. 1994, Nature,
368, 599
\item{  }Terasawa, N., \&  Sato, K. 1989a, Prog. Theor. Phys., 81, 254.
\item{  }Terasawa, N., \&  Sato, K. 1989b, Phys. Rev., D39, 2893.
\item{  }Terasawa, N., \&  Sato, K. 1989c, Prog. Theor. Phys., 81, 1085.
\item{ }Terasawa, N., \&  Sato, K., 1990, Ap. J. Lett., 362, L47.
\item{ }Thomas et al. 1994, ApJ, in press
\item{ }Trimble, V. 1987, Ann. Rev. of Astr., 25, pg.425, eds
\item{ }van den Bergh, A. 1989, Astr. Astroph. Rev., 1, 111
\item{ }Wagoner,  R.V., Fowler, W.A., \&  Hoyle, F. 1967, ApJ, 148, 3
\item{ }Wagoner, R. V. 1973, ApJ, 197, 343
\item{ }Walker, T.P., Steigman, G., Schramm, D.N., Olive, K.A., \& Kang, H.
1991, ApJ, 376, 51
\item{ }White, S.D.M., 1990, in {\it Physics of the Early Universe, Proc. 36th
Scottish Universities Summer School in Physics}, eds Peacock, J.A., Heavens,
A.F., \& Davies, A.T.
\item{ }Witten, E. 1984, Phys. Rev. D30, 272
\item{ }Wolfe, A.M. 1988, in {\it QSO Absorption Lines: Probing the Universe,
Proc. Space Telescope Science Institute Symp. No. 2}, pg.297, eds Blades, J.C.,
Turnshek, D., \& Norman, C.A., Cambridge University Press, Cambridge.
\item{ }Yang, J., Turner, M.S., Steigman, G., Schramm, D.N., \& Olive, K.A.
1984, ApJ, 281, 493

\vfill\eject

\centerline{\bf 6. Figure Captions}
\vskip 0.1in

\centerline{ }
\item{\bf Figure 1}The baryon-to-photon ratio $\eta$ as a function of length
coordinate $x$. We show a bimodal distribution with three high-density regions
at $\eta^h\approx 3\times 10^{-10}$ and low-density regions at $\eta^l<<3\times
10^{-10}$. The mean separation between centers of high-density regions is
denoted by $l$.

\centerline{ }
\item{\bf Figure 2}Nucleosynthesis yields resulting from a bimodal
baryon-to-photon distribution similar to the distribution shown in Figure 1. We
have assumed a regular lattice of spherically symmetric high-density regions
with step-function profiles and $\eta^h=3.1\times 10^{-10}$ embedded in a
low-density background with $\eta^l=3.1\times 10^{-15}$. We have taken a
fraction $f_V=0.065$ of the
cosmic volume to be filled with high-density regions, implying an average
${\bar\Omega_b}=0.003h_{50}^{-2}$. We show light-element abundance
 yields as a function of $l_{100}$ in meters, where $l_{100}$ is the proper
separation between centers of high-density regions at cosmic temperature
$T=100$ MeV.  The upper panel shows the $^4$He mass fraction $Y_p$, whereas
the center and lower panels show number fractions relative to hydrogen
for $^7$Li, and the sum of $^2$H and $^3$He, respectively. Observationally
inferred lower and
upper limits on the light-element abundances are taken from
Smith {\it et al.} (1993) and are indicated by the dotted boxes.

\end